\documentstyle[twocolumn,prc,aps,psfig]{revtex}
\begin{document}
\title{
\vspace*{-1.5em}
\hfill {\normalsize ADP-00-17/T401, UGI-00-6} \vspace{1em}\\ 
Strangeness production from $\pi N$ 
collisions in nuclear matter}

\author{K. Tsushima$^1$, A. Sibirtsev$^{1,2}$, A.W. Thomas$^1$}
\address{
$^1$Special Research Center for the Subatomic Structure of Matter 
and Department of Physics and Mathematical Physics \\
University of Adelaide, SA 5005, Australia \\
$^2$ Institut f\"ur Theoretische Physik, Universit\"at Giessen,
D-35392 Giessen, Germany}

\maketitle

\begin{abstract}
Kaon production in pion-nucleon collisions in nuclear matter
is studied in the resonance model. To evaluate the in-medium 
modification of the reaction amplitude as a function of the 
baryonic density we introduce relativistic, mean-field potentials 
for the initial, final 
and intermediate mesonic and baryonic states. These  vector and 
scalar potentials were calculated using 
the quark-meson coupling (QMC) model. 
The in-medium kaon production  cross sections in pion-nucleon interactions
for reaction channels with $\Lambda$ and $\Sigma$
hyperons in the final state were calculated at the  baryonic
densities appropriate to relativistic heavy ion collisions. 
Contrary to earlier work which has not allowed for the change of the
cross section in medium, we find that the data for kaon production
are consistent with a repulsive $K^+$-nucleus potential.
\end{abstract}


\section{Introduction}
The properties of kaons in nuclear matter have recently  attracted 
enormous interest because of their capacity to signal chiral symmetry 
restoration or give information on the possibility of kaon 
condensation in neutron 
stars~\cite{Kaplan,Brown1,Ko1,Lee}. Studies with a variety of 
models~\cite{Waas1,Waas2,Tsushima1,Sibirtsev1} indicate that the  
antikaon potential is attractive while the kaons feel a repulsive 
potential in nuclear matter. The results from kaonic 
atoms~\cite{Friedman1,Friedman2},  as well as an
analysis~\cite{Li1,Cassing2,Li2,Bratkovskaya,Cassing3} of the $K^-$ 
production from heavy ion 
collisions~\cite{Schroter,Senger1,Barth,Laue,Senger2},
are in reasonable agreement with the former expectation for
antikaons. However, the analysis of  available data on 
$K^+$ production from heavy ion collisions at SIS 
energies~\cite{Barth,Laue,Senger2,Senger3} contradicts 
the predictions that the kaon potential is repulsive. 
The comparison between the heavy ion calculations and the 
data~\cite{Li2,Bratkovskaya,Cassing3,Senger3,Li3} indicates
that the $K^+$-meson spectra are best described by neglecting  
any in-medium modification of the kaon properties. Furthermore, the 
introduction of even a weakly repulsive 
$K^+$-nucleus potential results in a 
substantial underestimate of the experimental data on kaon 
production. 

Since in  heavy ion collisions at
SIS energies~\cite{Schroter,Senger1,Barth,Laue,Senger2}  
the $K^+$-mesons are predominantly produced by secondary pions,
we investigate the kaon production reactions, 
$\pi{+}N{\to}Y{+}K$ ($Y = \Lambda, \Sigma$ hyperons), 
in nuclear matter. 
To be specific, we combine earlier studies of kaon production
in free space with a very successful, relativistic mean field
description of nuclear systems (QMC). All parameters are fixed by
earlier studies and the effects of the medium on the reaction cross
sections are calculated for the first time. The result is impressive in
that the medium effects explain the nuclear production data, without any
adjustment of the parameters determined elsewhere, {\em including} the
standard repulsive kaon-nucleus interaction.

Our paper is organized as follows. In Sec.~II we introduce the
vector and scalar potentials for mesons and baryons involved in
the calculations of the $\pi{+}N{\to}Y{+}K$ amplitudes. 
We explain in Sec.~III the resonance model which is used to 
calculate the cross sections, $\pi{+}N{\to}Y{+}K$. 
The strangeness production threshold in nuclear matter and its dependence 
on the baryon density is discussed in Sec.~IV. The
cross sections of kaon production in $\pi{+}N$
collisions in vacuum and in nuclear matter at different densities
are then evaluated and shown in Sec.~V for 
the $\pi^-{+}p{\to}\Lambda{+}K^0$ reaction
and in Sec.~VI for the $\pi{+}N{\to}\Sigma{+}K$
reactions. The impact of our results on heavy ion collisions
is discussed in Sec.~VII. Finally, the summary and conclusions are given 
in Sec.~VIII. 

\section{Mean-field potentials for mesons and baryons}
In the present study, we use the quark-meson coupling (QMC) 
model~\cite{Guichon},
which has been successfully applied not only to the problems
of conventional nuclear physics~\cite{Guichonf,Saito2}
but also to the studies of meson and hyperon properties in a nuclear
medium~\cite{Tsushima1,Tsushimak1,Tsushimak2,Tsushimak3,Tsushimak4,Tsushimak5,Saitok1,Saitok2,Tony,Hyperon1,Hyperon2}.
A detailed description of the Lagrangian density and the
mean-field equations
are given in
Refs.~\cite{Tsushima1,Guichonf,Saito2,Tsushimak1,Tsushimak2,Tsushimak3,Saitok1}.
The Dirac equations for the quarks and antiquarks in the hadron bags 
($q = u,\bar{u},d$ or $\bar{d}$, hereafter), 
neglecting the Coulomb force,
are given by:
\begin{eqnarray}
\left[ i \gamma \cdot \partial_x -
(m_q - V^q_\sigma)
\mp \gamma^0
\left( V^q_\omega +
\frac{1}{2} V^q_\rho
\right) \right] \nonumber \\
\times \left( \begin{array}{c} \psi_u(x)  \\
\psi_{\bar{u}}(x) \\ \end{array} \right)
 = 0,
\label{diracu}
\\
\left[ i \gamma \cdot \partial_x -
(m_q - V^q_\sigma)
\mp \gamma^0
\left( V^q_\omega
- \frac{1}{2} V^q_\rho
\right) \right] \nonumber \\  \times
\left(\begin{array}{c} \psi_d(x) \\ \psi_{\bar{d}}(x) \\ \end{array} \right)
 = 0,
\label{diracd}
\\
\left[ i \gamma \cdot \partial_x - m_{s} \right]
\psi_{s} (x)\,\, ({\rm or}\,\, \psi_{\bar{s}}(x)) = 0.
\label{diracsc}
\end{eqnarray}
The mean-field potentials for a bag in  nuclear matter
are defined by $V^q_\sigma{=}g^q_\sigma
\sigma$,
$V^q_\omega{=}$ $g^q_\omega
\omega$ and
$V^q_\rho{=}g^q_\rho b$,
with $g^q_\sigma$, $g^q_\omega$ and
$g^q_\rho$ the corresponding quark-meson coupling
constants. 

The normalized, static solution for the ground state quarks or antiquarks
in the hadron, $h$, may be written 
as~\cite{Tsushima1,Tsushimak1,Tsushimak2,Tsushimak3}:
\begin{eqnarray}
\psi_f (x) = N_f e^{- i \epsilon_f t / R_h^*}
\psi_f (\mbox{\boldmath $x$}),
\label{wavefunction}
\end{eqnarray}
where $f$ labels the
quark flavours, and $N_f$ and $\psi_f(\mbox{\boldmath $x$})$
are the normalization factor and
corresponding spin and spatial part of the wave function. The bag
radius in medium, $R_h^*$, which generally depends on the hadron species to
which the quarks and antiquarks belong, will be determined through the
stability condition for the (in-medium) mass of the meson against the
variation of the bag 
radius~\cite{Tsushima1,Guichonf,Saito2,Tsushimak1,Tsushimak2,Saitok1}
(see also Eq.~(\ref{equil})). The eigenenergies, $\epsilon_f$, in
Eq.~(\ref{wavefunction}) in units of $1/R_h^*$, are given by
\begin{eqnarray}
\left( \begin{array}{c}
\epsilon_u \\
\epsilon_{\bar{u}}
\end{array} \right)
&=& \Omega_q^* \pm R_h^* \left(
V^q_\omega
+ \frac{1}{2} V^q_\rho \right),
\label{uenergy}
\\
\left( \begin{array}{c} \epsilon_d \\
\epsilon_{\bar{d}}
\end{array} \right)
&=& \Omega_q^* \pm R_h^* \left(
V^q_\omega
- \frac{1}{2} V^q_\rho \right),
\label{denergy}
\\
\epsilon_{s}
&=& \epsilon_{\bar{s}} =
\Omega_{s},
\label{cenergy}
\end{eqnarray}
where $\Omega_q^*
{=}\sqrt{x_q^2{+}(R_h^* m_q^*)^2}$, with
$m_q^*{=}m_q{-}g^q_\sigma \sigma$ and
$\Omega_{s}{=}\sqrt{x_{s}^2{+}(R_h^* m_{s})^2}$.
The bag eigenfrequencies, $x_q$ and $x_{s}$, are
determined by the usual, linear boundary condition~\cite{Guichonf,Saito2}.
Note that the lowest eigenenergy  for the Dirac equation
(Hamiltonian) for the quark, which is positive, can be regarded 
as the analog of a constituent quark mass.

The hadron masses
in symmetric nuclear matter relevant for the present study 
are calculated by:
\begin{eqnarray}
m_h^* &=& \frac{(n_q + n_{\bar{q}}) \Omega_q^*
+ (n_s + n_{\bar{s}}) \Omega_s - z_h}{R_h^*}
+ {4\over 3}\pi R_h^{* 3} B,
\label{md}
\\
& &\left. \frac{\partial m_h^*}
{\partial R_h}\right|_{R_h = R_h^*} = 0.
\label{equil}
\end{eqnarray}
In Eq.~(\ref{md}), $n_q$ ($n_{\bar{q}}$) and $n_s$ ($n_{\bar{s}}$)
are the lowest mode light quark (antiquark) and strange 
(antistrange) quark numbers in the hadron, $h$, respectively,
and the $z_h$ parametrize the sum of the
center-of-mass and gluon fluctuation effects, and are assumed to be
independent of density. The parameters are determined in free space to
reproduce their physical masses.

\begin{figure}[htb]
\vspace{-2mm}
\psfig{file=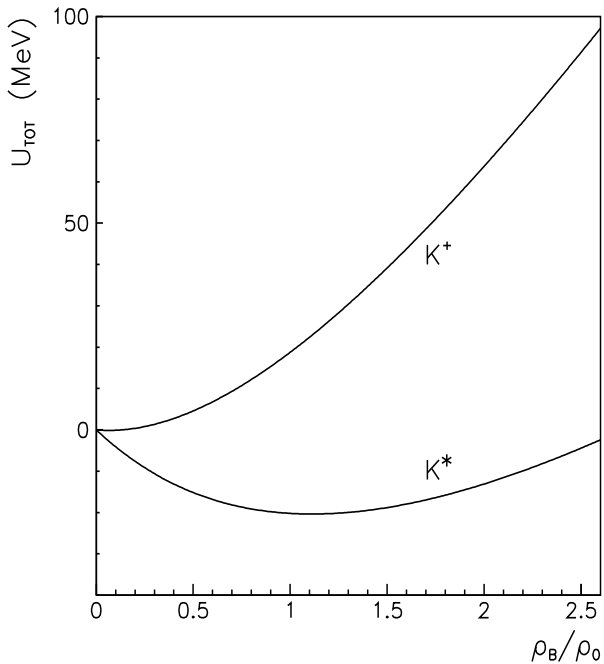,width=9cm,height=10.cm}
\caption[]{\label{laka9}Total potential $U_{tot}$ for $K^+$  and 
$K^\ast(892)^+$ mesons plotted as function of the baryon density, $\rho_B$, 
in units of  saturation density, $\rho_0$=0.15 
fm$^{-3}$ of the nuclear matter.}
\vspace{5mm}
\end{figure}

In this study we chose the values $m_q$=5 MeV and 
$m_s$=250 MeV for the current quark masses, and $R_N{=}0.8$
fm for the bag radius of the nucleon in free space. Other input
parameters and some of the quantities calculated are given in 
Refs.~\cite{Guichonf,Saito2,Tsushimak1,Tsushimak2,Tsushimak3}.
We stress that while the model has a
number of parameters, only three of them, $g^q_\sigma$, $g^q_\omega$
and $g^q_\rho$, are adjusted to fit nuclear data -- namely the
saturation energy and density of symmetric nuclear matter and the bulk
symmetry energy. None of the results for nuclear properties depend
strongly on the choice of the other parameters -- for example, the
relatively weak dependence of the final results for the properties of
finite nuclei, on the chosen values of the current quark mass and bag
radius, is shown explicitly in Refs.~\cite{Guichonf,Saito2}. Exactly
the same coupling constants, $g^q_\sigma$, $g^q_\omega$ and
$g^q_\rho$, are used for the light quarks in the mesons and hyperons as in 
the nucleon. 
However, in studies of the kaon system, we found that it was
phenomenologically necessary to increase the strength of the vector
coupling to the non-strange quarks in the $K^+$ (by a factor of
$1.4^2$) in order to reproduce the empirically extracted $K^+$-nucleus
interaction~\cite{Waas2,Tsushima1,Sibirtsev1,Waas3,likpot}, which is 
slightly repulsive if one wants to be consistent with the $K^+ N$ scattering
length, and the corresponding value at $\rho_B = 0.16$ fm$^{-3}$ is
estimated to be about 20 MeV~\cite{likpot}. 
Thus, we will use the stronger vector potential,
$1.4^2 V^q_\omega$, for the $K^+$-meson in this study.   
Calculated mean field potential felt by $K^+$-meson, using 
$1.4^2 V^q_\omega$, is shown in FIG.~\ref{laka9}.
Through Eqs.~(\ref{diracu}) -- (\ref{equil}) and usual QMC 
formalism~\cite{Tsushima1,Guichonf,Saito2,Tsushimak1,Tsushimak2,Tsushimak3,Saitok1} we self-consistently
calculate effective masses,
$m^*_h$,
and mean field potentials, $V^q_{\sigma,\omega,\rho}
$, in symmetric nuclear matter.
The scalar ($U^{h}_s$) and vector ($U^{h}_v$) potentials 
felt by the hadrons, $h$,  
in nuclear matter are given by:
\begin{eqnarray}
U^{h}_s
&\equiv& U_s = m^*_h - m_h,
\label{spot}\\
U^{h}_v &=&
  (n_q - n_{\bar{q}}) {V}^q_\omega - I_3 V^q_\rho, 
\,\,(V^q_\omega \to 1.4^2 {V}^q_\omega\,\, {\rm for}\, K^+). 
\label{vdpot}
\end{eqnarray}
%

\begin{figure}[htb]
\vspace{-4mm}
\psfig{file=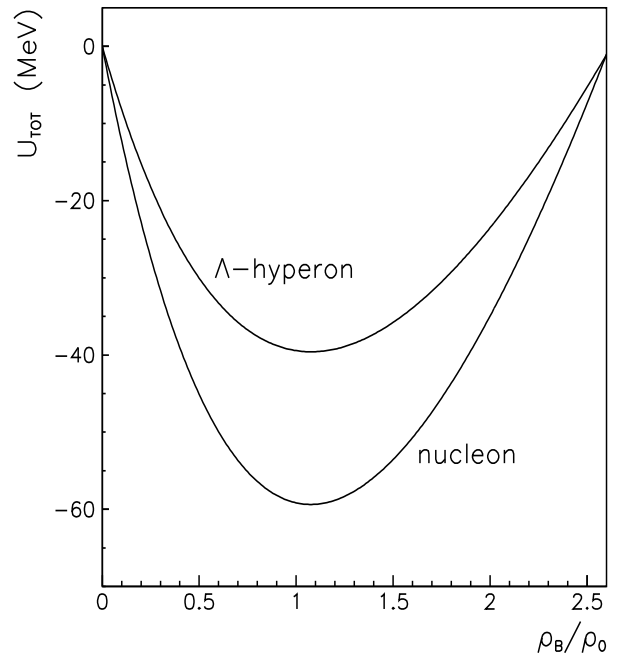,width=9cm,height=10.cm}
\caption[]{\label{laka8}Total potential $U_{tot}$ for nucleon and
$\Lambda$-hyperon shown as a function of the baryon density, $\rho_B$,
in units of the nuclear matter saturation density, $\rho_0$=0.15
fm$^{-3}$.}
\vspace{5mm}
\end{figure}

In Eq.~(\ref{vdpot}), $I_3$ is the third component of isospin projection  
of the hadron, $h$, and the $\rho$ meson mean field potential, $V^q_\rho$,
is zero in symmetric nuclear matter.
Then, within the approximation that the mean field potentials are 
independent of momentum, 
the four-momentum of the hadron is modified by, 
$p^\mu_h = (\sqrt{{\mbox{\boldmath $p$}}^2 + m_h^{*2}} + U^{h}_v, 
{\mbox{\boldmath $p$}}$), which 
modifies not only the kinematical factors such as the flux and 
the phase space, but also modifies the reaction amplitudes.
Obviously, the reaction thresholds are modified in nuclear matter 
and now depend on the baryon density.

FIG.~\ref{laka8} shows the total ($U_{tot}$) nucleon and $\Lambda$-hyperon
potentials at zero momenta as function of the baryon density, in
units of the saturation density of nuclear matter $\rho_0$=0.15 
fm$^{-3}$. Let us recall that at momentum {\mbox{\boldmath $p$}} $= 0$, 
\begin{equation}
U_{tot}= U_s + U_v,
\end{equation}
and the total potential for the $\Sigma$-hyperon is almost equal
to that for the $\Lambda$-hyperon.

FIG.~\ref{laka8} indicates that both nucleon and hyperon
potentials approach minima around normal nuclear matter density,
which reflects the fact that around $\rho_0$ the energy density 
of nuclear matter is minimized. 
FIG.~\ref{laka9} shows the density dependence of the total
$K$ and $K^\ast$-meson potentials at zero momenta.
The total kaon potential is repulsive as explained before, 
and depends substantially on
the baryon density. The $K^\ast$-meson total potential is attractive
at baryon densities below $\simeq$2.7$\rho_0$.

Now, we will discuss the in-medium modification of the resonance masses.
At present it seems that there is no reliable estimate for the in-medium 
modification of masses for the higher mass baryon resonances.  
In view of its numerous successful applications elsewhere, we base
our estimate on the QMC model~\cite{Hyperon1,Hyperon2}.

We assume that the light quarks in the baryon resonances are   
responsible for the mass modification in nuclear medium,   
as in the QMC model~\cite{Hyperon1,Hyperon2}.
However, there is a possibility that the excited state light quarks 
may couple differently to the scalar $\sigma$ field from those in the 
ground states, although we expect the difference is small.
Thus, we estimate the range for the in-medium baryon resonance masses  
by the following two extreme cases, i.e., (i) all light quarks 
including those in the excited states play the same role for the 
mass modification as those in the ground states,
(ii) only the ground state light quarks play the role 
as in the usual QMC model.
These two cases are expected to give the maximum 
and minimum limits for the mass modifications 
of the baryon resonances. 
Specifically, the range for the effective masses  
of the baryon resonance in medium is given   
(see, e.g., Ref.~\cite{Bhaduri} for the quark model basis of the 
baryon resonances) :

\begin{eqnarray}
m_{N(1650)} - \delta m^*_N &\leq& m^*_{N(1650)} \leq 
m_{N(1650)} - \frac{2}{3} \delta m^*_N, \label{mr1} \\
m_{N(1710)} - \delta m^*_N &\leq& m^*_{N(1710)} \leq
m_{N(1710)} - \frac{1}{3} \delta m^*_N, \label{mr2} \\
m_{N(1720)} - \delta m^*_N &\leq& m^*_{N(1720)} \leq
m_{N(1720)} - \frac{1}{3} \delta m^*_N, \label{mr3} \\
m_{\Delta(1920)} - \delta m^*_N &\leq& m^*_{\Delta(1920)} \leq  
m_{\Delta(1920)} - \frac{1}{3} \delta m^*_N, \label{mr4} \\
{\rm with} \hspace{2em} \delta m^*_N &=& m_N - m^*_N. \label{dmn}
\label{mresonance}
\end{eqnarray}

These in-medium resonance masses may be expected to modify the 
resonance propagators in the reaction amplitudes.
To avoid introducing extra unknown parameters in this initial
study, we approximate  the in-medium resonance widths appearing 
in the propagator by the free space ones. 
From Eqs.~(\ref{mr1}) --~(\ref{dmn}), we will show results for the 
cross section calculated using the lower limit for the resonance masses. 
However, we have also performed the calculation using the upper limit 
for the resonance masses, and confirmed that our conclusion 
remains the same.

\section{Resonance model}
We explain in this section the resonance model 
~\cite{Tsushima2,Tsushima3,Tsushima3a,Tsushima4,Tsushima5,Sibirtsev2,Sibirtsev3}, 
which could describe the energy dependence of the total cross sections, 
$\pi N \to Y K$, quite well, and has been used widely in 
kaon production simulation 
codes~\cite{Cassing2,Li2,Bratkovskaya,Li3,likpot,kcode}.
We extend the model by including medium modification 
of the hadron properties, not only in the kinematic factors such as 
the flux and the phase space, but also in the reaction amplitudes.

\begin{figure}[htb]
\vspace*{-2.5cm}
\hspace*{-1.5cm}
\psfig{file=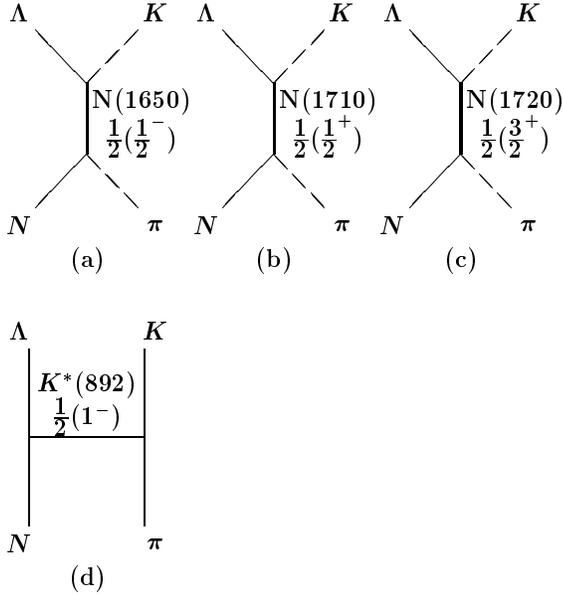,width=9.5cm,height=14.cm}
\vspace{-3cm}
\caption[]{\label{pbyklafig}$K$ and $\Lambda$ production processes.}
\end{figure}

\begin{figure}[htb]
\vspace*{-2.5cm}
\hspace*{-1.5cm}
\psfig{file=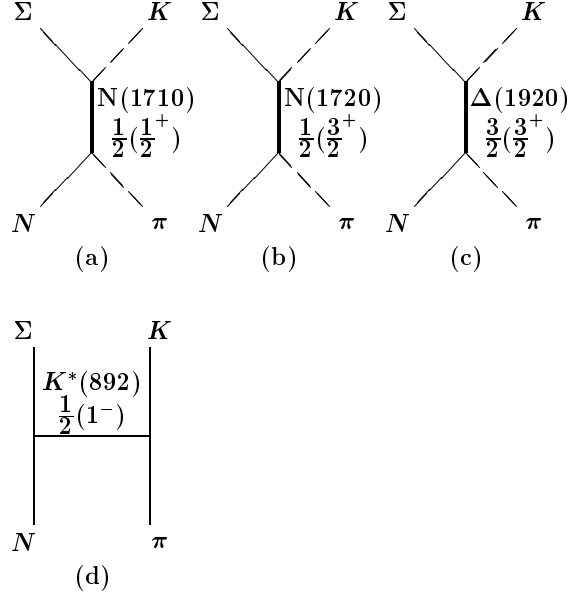,width=9.5cm,height=14.cm}
\vspace{-3cm}
\caption[]{\label{pbyksifig}$K$ and $\Sigma$ production processes.}
\end{figure}

We consider kaon and hyperon production processes in $\pi N$ collisions 
shown in FIGs.~\ref{pbyklafig} and~\ref{pbyksifig}.
Because different intermediate states and final states 
contribute to the $\pi N \to \Lambda K$ and $\pi N \to \Sigma K$ reactions, 
the in-medium modification of the reaction amplitudes is also expected to be 
different, as will indeed be shown later.

In TABLE~\ref{resonance} we summarize the data for the
resonances which are included in the model.

   
The effective Lagrangian densities used for evaluating the 
processes shown in FIGs.~\ref{pbyklafig} and~\ref{pbyksifig} are:
\begin{eqnarray}
{\cal L}_{\pi N N} &=& 
-ig_{\pi N N} \bar{N} \gamma_5 \vec\tau N \cdot \vec\pi,\label{pnn}\\
{\cal L}_{\pi N N(1650)} &=&-g_{\pi N N(1650)}\nonumber\\
&&\hspace*{-5em}\times 
\left( \bar{N}(1650) \vec\tau N \cdot \vec\pi
+ \bar{N} \vec\tau N(1650) \cdot \vec\pi\,\, \right),
\label{pna}\\
{\cal L}_{\pi N N(1710)} &=&-ig_{\pi N N(1710)}\nonumber\\
&&\hspace*{-5em}\times
\left( \bar{N}(1710) \gamma_5 \vec\tau N \cdot \vec\pi
+ \bar{N} \gamma_5 \vec\tau N(1710) \cdot \vec\pi\,\, \right),
\label{pnb}\\
{\cal L}_{\pi N N(1720)} &=&\frac{g_{\pi N N(1720)}}{m_\pi}\nonumber\\ 
&&\hspace*{-5em}\times
\left( \bar{N}^\mu(1720) \vec\tau N \cdot \partial_\mu \vec\pi
+ \bar{N} \vec\tau N^\mu(1720) \cdot \partial_\mu \vec\pi \, \right),
\label{pnc}\\
{\cal L}_{\pi N \Delta(1920)} 
&=&\frac{g_{\pi N \Delta(1920)}}{m_\pi}\nonumber\\
&&\hspace*{-7em}\times
\left( \bar{\Delta}^\mu(1920) \overrightarrow{\cal I} N \cdot
\partial_\mu \vec\pi + \bar{N} {\overrightarrow{\cal I}}^\dagger
\Delta^\mu(1920) \cdot \partial_\mu \vec\pi \, \right),
\label{pnd}\\
{\cal L}_{K \Lambda N(1650)} &=&-g_{K \Lambda N(1650)}\nonumber\\
&&\hspace*{-5em}\times
\left( \bar{N}(1650) \Lambda K + \bar{K} \bar\Lambda  N(1650) \right),
\label{kla}\\
{\cal L}_{K \Lambda N(1710)} &=&-ig_{K \Lambda N(1710)}\nonumber\\
&&\hspace*{-5em}\times
\left( \bar{N}(1710) \gamma_5 \Lambda K
+ \bar{K} \bar\Lambda \gamma_5 N(1710) \right),
\label{klb}\\
{\cal L}_{K \Lambda N(1720)} &=&\frac{g_{K \Lambda N(1720)}}{m_K}\nonumber\\
&&\hspace*{-5em}\times
\left( \bar{N}^\mu(1720) \Lambda
\partial_\mu K + (\partial_\mu \bar{K}) \bar\Lambda N^\mu(1720) \right),
\label{klc}\\
{\cal L}_{K \Sigma N(1710)} &=&-ig_{K \Sigma N(1710)}\nonumber\\
&&\hspace*{-5em}\times
\left( \bar{N}(1710) \gamma_5 \vec\tau \cdot \overrightarrow\Sigma K
+ \bar{K} \overrightarrow{\bar \Sigma} \cdot \vec\tau
\gamma_5 N(1710) \right),
\label{ksb}\\
{\cal L}_{K \Sigma N(1720)} &=&\frac{g_{K \Sigma N(1720)}}{m_K}\nonumber\\
&&\hspace*{-7em}\times
\left( \bar{N}^\mu(1720) \vec\tau \cdot \overrightarrow\Sigma
\partial_\mu K + (\partial_\mu \bar{K}) \overrightarrow{\bar \Sigma}
\cdot \vec\tau N^\mu(1720) \right),
\label{ksc}\\
{\cal L}_{K \Sigma \Delta(1920)} 
&=&\frac{g_{K \Sigma \Delta(1920)}}{m_K}\nonumber\\
&&\hspace*{-7.5em}\times
\left( \bar{\Delta}^\mu(1920) \overrightarrow{\cal I}
\cdot \overrightarrow\Sigma \partial_\mu K
+ (\partial_\mu \bar{K}) \overrightarrow{\bar \Sigma} \cdot
{\overrightarrow{\cal I}}^\dagger \Delta^\mu(1920) \right),
\label{ksd}\\
{\cal L}_{K^*(892) K \pi} &=&ig_{K^*(892) K \pi}\nonumber\\ 
&&\hspace*{-7.5em}\times
\left( \bar{K} \vec\tau K^*_\mu(892) \cdot
\partial^\mu \vec\phi
- (\partial^\mu \bar{K}) \vec\tau K^*_\mu(892) \cdot \vec\phi \,\right)
+{\rm h. c.},
\label{kskpi}\\
{\cal L}_{K^*(892)  \Lambda  N} &=&-g_{K^*(892)  \Lambda N }\nonumber\\
&&\times
\left( \bar{N} \gamma^\mu  \Lambda K^*_\mu(892)
+ {\rm h. c.} \right),
\label{kslan}\\
{\cal L}_{K^*(892) \Sigma N} &=&-g_{K^*(892) \Sigma N}\nonumber\\
&&\times
\left( \bar{N} \gamma^\mu \vec\tau \cdot \overrightarrow\Sigma K^*_\mu(892)
+ {\rm h. c.} \right).
\label{kssin}
\end{eqnarray}
In the above, the operators $\overrightarrow{\cal I}$
and $\overrightarrow{\cal K}$ are defined by
\begin{eqnarray}
\overrightarrow{\cal I}_{M\mu} &\equiv&  \displaystyle{\sum_{\ell=\pm1,0}}
(1 \ell \frac{1}{2} \mu | \frac{3}{2} M)
\hat{e}^*_{\ell},\\
\overrightarrow{\cal K}_{M M'} &\equiv&  \displaystyle{\sum_{\ell=\pm1,0}}
(1 \ell \frac{3}{2} M' | \frac{3}{2} M)
\hat{e}^*_{\ell},\label{ccdelta}
\end{eqnarray}
with $M$, $\mu$ and $M'$ being the third components of the
isospin projections,
and $\vec \tau$ the Pauli matrices.
$N, N(1710), N(1720)$ and $\Delta(1920)$
stand for the fields of the nucleon
and the corresponding baryon resonances. They are expressed by
$\bar{N} = \left( \bar{p}, \bar{n} \right)$,
similarly for the nucleon resonances, and
$\bar{\Delta}(1920) = ( \bar{\Delta}(1920)^{++},
\bar{\Delta}(1920)^+, \bar{\Delta}(1920)^0,
\bar{\Delta}(1920)^- )$ in isospin space.
The physical representations of the kaon field are,
$K^T = \left( K^+, K^0 \right)$ and
$\bar{K} = \left( K^-, \bar{K^0} \right)$, respectively,
and similarly for the $K^*(892)$, where the superscript $T$ means the
transposition operation.
They are defined as annihilating (creating) the physical particle
(anti-particle) states.
For the propagators $iS_F(p)$ of the spin 1/2 and
$iG^{\mu \nu}(p)$ of the spin 3/2 resonances we use:
\begin{equation}
iS_F(p) = i \frac{\gamma \cdot p + m}{p^2 - m^2 + im\Gamma^{full}}\,,
\label{spin1/2}
\end{equation}
\begin{equation}
iG^{\mu \nu}(p) = i \frac{-P^{\mu \nu}(p)}{p^2 - m^2 +
im\Gamma^{full}}\,,  \label{spin3/2}
\end{equation}
with
\begin{eqnarray}
P^{\mu \nu}(p) &=& - (\gamma \cdot p + m)\nonumber\\
&&\hspace*{-5em}\times
\left[ g^{\mu \nu} - \frac{1}{3} \gamma^\mu \gamma^\nu
- \frac{1}{3 m}( \gamma^\mu p^\nu - \gamma^\nu p^\mu)
- \frac{2}{3 m^2} p^\mu p^\nu \right], \label{pmunu}
\end{eqnarray}
where $m$ and $\Gamma^{full}$ stand for the mass and full decay
width of the corresponding resonances.
For the form factors, $F (\vec{q})$
($\vec{q}$ is the momentum of meson, $\pi$ or $K$),
appearing in the meson-baryon-(baryon resonance) vertices, we use:
\begin{equation}
F(\vec{q}) =
\displaystyle{\left( \frac{\Lambda^2}{\Lambda^2
+ \vec{q}\,^2} \right)}.
\label{form1}
\end{equation}
For the $K^*(892)$-$K$-$\pi$ vertex  we adopt the form
factor of Ref.~\cite{gob}:
\begin{eqnarray}
F_{K^*(892) K \pi}(|\frac{1}{2}(\vec{p}_K-\vec{p}_\pi)|)
&=&\nonumber\\
&&\hspace{-12em}\times
C |\frac{1}{2}(\vec{p}_K-\vec{p}_\pi)|
\exp\left( - \beta |\frac{1}{2}(\vec{p}_K-\vec{p}_\pi)|^2 \right).
\label{form2}
\end{eqnarray}

In the calculation, the form factors of
Eqs. (\ref{form1}) and (\ref{form2}) are multiplied by the corresponding
coupling constants.

\begin{table}[htb]
\caption{\label{resonance}
Resonances included in the model.
Confidence levels of the resonances are,
$N(1650)****$, $N(1710)***$, $N(1720)****$ and
$\Delta(1920)***$~\protect\cite{PDG}. Note that
the $\Delta(1920)$ resonance is treated as an effective
resonance which effectively represents the contributions of six resonances,
$\Delta(1900)$, $\Delta(1905)$, $\Delta(1910)$, $\Delta(1920)$,
$\Delta(1930)$ and $\Delta(1940)$.
See Refs.~\protect\cite{Tsushima2,Tsushima3a} for this effective treatment
of the $\Delta(1920)$.}
\begin{center}
\begin{tabular}{c|cccc}
\hline
Resonance &Width &Channel &B.R. ratio &Used\\
\hline
N(1650)$\,(\frac{1}{2}^-)$ &150   &$N \pi$      &0.60 -- 0.80 &0.700 \\
                           &(MeV) &$\Lambda K$  &0.03 -- 0.11 &0.070 \\
\hline
N(1710)$\,(\frac{1}{2}^+)$ &100   &$N \pi$      &0.10 -- 0.20 &0.150 \\
                           &(MeV) &$\Lambda K$  &0.05 -- 0.25 &0.150 \\
                           &      &$\Sigma K$   &0.02 -- 0.10 &0.060 \\
\hline
N(1720)$\,(\frac{3}{2}^+)$ &150   &$N \pi$      &0.10 -- 0.20 &0.150 \\
                           &(MeV) &$\Lambda K$  &0.03 -- 0.10 &0.065 \\
                           &      &$\Sigma K$   &0.02 -- 0.05 &0.035 \\
\hline
$\Delta$(1920)$\,(\frac{3}{2}^+)$ &200 &$N \pi$ &0.05 -- 0.20 &0.125 \\
                           &(MeV)      &$\Sigma K$   &0.01 -- 0.03 &0.020 \\
\hline
$K^*(892)\,(1^-)$ &50 &$K \pi$ &$\sim$ 1.00&1.00\\
                  &(MeV)       &    & \\
\hline
\end{tabular}
\end{center}
\end{table}

The parameters of the model, namely the form factors
in the interaction vertices and the coupling constants, were
fixed by available experimental data on the different
$\pi{+}N{\to}Y{+}K$ reaction channels. Furthermore, the same parameters
which were determined by the $\pi{+}N{\to}Y{+}K$ reactions
were used in the calculations of strangeness
production in baryon-baryon collisions, and they also
reproduced the available data reasonably
well~\cite{Tsushima5,Sibirtsev2,Sibirtsev3}.
In TABLE~\ref{cconst} we list all values for the coupling constants, 
cut-offs, $C$ and $\beta$ parameters relevant for the present study. 

\begin{table}[htb]
\caption{\label{cconst}
Values for coupling constants, cut-offs, $C$ and $\beta$ 
parameters used in the present study.
For details about the coupling
constants relevant for $\Delta(1920)$, $g_{\pi N \Delta(1920)}$ and
$g_{K \Sigma \Delta(1920)}$, 
see Refs.~\protect\cite{Tsushima2,Tsushima3a}.
}
\begin{center}
\begin{tabular}{c|c|c}
\hline
vertex & $g^2/4\pi$ & cut-off $\Lambda$ (MeV)\\
\hline
 & & \\
 $\pi N N$  & $14.4$ & $1050$\\
 $\pi N N(1650)$ & $1.12 \times 10^{-1}$ &$800$\\
 $\pi N N(1710)$ &$2.05 \times 10^{-1}$ &$800$\\
 $\pi N N(1720)$ &$4.13 \times 10^{-3}$ &$800$\\
 $\pi N \Delta(1920)$ &$1.13 \times 10^{-1}$ &$500$\\
 $K \Lambda N(1650)$ &$5.10 \times 10^{-2}$ &$800$\\
 $K \Lambda N(1710)$ &$3.78$ &$800$\\
 $K \Lambda N(1720)$ &$3.12 \times 10^{-1}$ &$800$\\
 $K \Sigma N(1710)$ &$4.66$ &$800$\\
 $K \Sigma N(1720)$ &$2.99 \times 10^{-1}$ &$800$\\
 $K \Sigma \Delta(1920)$ & $3.08 \times 10^{-1}$ & $500$\\
 $K^*(892) \Lambda N$ & $1.62 \times 10^{-2}$ & $1200$\\
 $K^*(892) \Sigma  N$ & $1.62 \times 10^{-2}$ & $1200$\\
\hline
 $K^*(892) K \pi$ &$5.48 \times 10^{-2}$ &$C=2.72 fm$\\
                  &                      &$\beta=8.88\times 10^{-3} fm^2$\\
\hline
\end{tabular}
\end{center}
\end{table}

Since the strength of the $\Lambda K$ and $\Sigma K$ coupling to 
the various baryonic resonances are different (see TABLE~1),
the dynamics of $\Lambda$ and $\Sigma$ production are 
also different. This can be understood as follows.

The amplitude of the resonance propagator in the reaction amplitude,    
$\pi N \to Y K$, 
becomes maximal when the invariant collision energy 
crosses the mass of the resonance.
Furthermore, the cross section $\sigma(\pi N \to Y K)$ is proportional to 
$(q_Y)^{2\ell + 1}$ near threshold (with $\ell$ and $q_Y = |\vec{q}_Y|$ 
the orbital angular momentum and momentum of the $Y K$ pair in the 
center of mass system). In seeking to understand the difference 
in the behaviour of $\sigma(\pi N \to \Lambda K)$ and 
$\sigma(\pi N \to \Sigma K)$ it is important to note that the former 
receives a large contribution from the $s$-wave $N(1650)$ (with $\ell = 0$),
while the latter is dominated by $p$-wave resonances (with $\ell = 1$).
The subtle interplay between the change in momentum dependence 
associated with threshold variations and the change in resonance 
amplitudes associated with varying resonance masses is responsible for the 
different behaviour of $\sigma(\pi N \to \Lambda K)$ and
$\sigma(\pi N \to \Sigma K)$.

The total free mass of the $\Sigma K$ system, $m_\Sigma +m_K$,
is very close to the mass of the $P_{11}(1710)$ resonance, i.e.,  
$M_{P_{11}}-m_\Sigma -m_K \simeq$27~MeV,
which dictates its  substantial role in $\Sigma$ hyperon
production~\cite{Sibirtsev3}. The situation is
different for the $\Lambda K$ system, since the nearest
baryonic resonance is $S_{11}(1650)$ and
$M_{S_{11}}- m_\Lambda -m_K\simeq$41~MeV. Furthermore,
because $S_{11}(1650)$ and $P_{11}(1710)$ have different
couplings to the final states, $\Lambda K$ (relative $s$-wave) 
and $\Sigma K$ (relative $p$-wave), respectively, 
the change in the vertices, $S_{11}(1650) \Lambda K$ and 
$P_{11}(1710) \Sigma K$ due to the momentum modified 
in medium are also different.
Furthermore, when threshold changes as the baryon desnsity varies, 
the change in the strength of the contributions from 
the $S_{11}(1650)$ to the 
$\pi N \to \Lambda K$ reaction 
and that of the $P_{33}(1920)$ to the $\pi N \to \Sigma K$ reaction  
should be different as a function of baryonic density. 
Thus, combining these effects 
for the $\pi N \to \Lambda K$ and $\pi N \to \Sigma K$ reactions, 
we can see that the energy dependence of the cross sections 
in a nuclear medium can be significantly modified. 

\section{The $\pi{+}N{\to}Y{+}K^+$ reaction threshold in nuclear matter}
The dispersion relation in nuclear matter relating the total energy 
$E$ and the momentum {\mbox{\boldmath $p$} of the particle is written as 
\begin{equation}
E=\sqrt{{\mbox{\boldmath $p$}}^2 + (m + U_s)^2 } + U_v, 
\label{dispa}
\end{equation}
where $m$ is the bare mass and $U_s$, $U_v$ denote the scalar and 
vector parts of the potential in nuclear matter. The threshold,  
$\sqrt{s_{th}}$, for the  reaction $\pi{+}N{\to}Y{+}K^+$ is given 
as the sum of the total energies of the final $K^+$-meson and 
$Y$-hyperon, taking their momenta to be zero and hence 
\begin{equation}
\sqrt{s_{th}}=m^K+m^Y+U_s^K+U_s^Y+U_v^K+U_v^Y, 
\label{th1}
\end{equation}
where now the upper indices denote kaons and hyperons. 
The solid lines in FIG.~\ref{laka10}
show the $K^+\Lambda$ and $K^+\Sigma$ reaction thresholds,
$\sqrt{s_{th}}$,  as a function of the baryon density.
Obviously, in free space the scalar and vector potential vanish
and the reaction threshold equals to the sum of 
the bare masses of the produced particles, which is shown by the
dashed lines in FIG.~\ref{laka10} for the $K^+\Lambda$ and $K^+\Sigma$ 
final states.

\begin{figure}[htb]
\vspace{-2mm}
\psfig{file=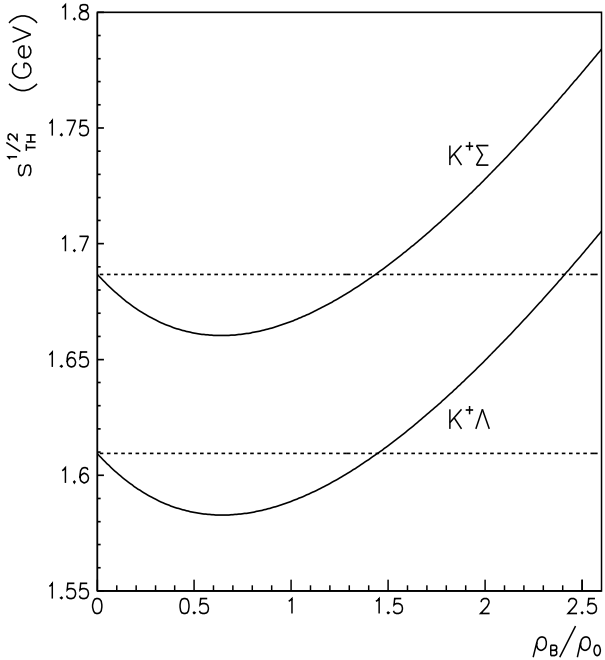,width=9cm,height=10.cm}
\caption[]{\label{laka10}The threshold energy, $\sqrt{s_{th}}$,
for $K^+\Lambda$
and $K^+\Sigma$ production given by their total in-medium energy
at zero momentum, as a function of the baryon density, $\rho_B$,
in units of saturation density of nuclear
matter, $\rho_0$=0.15 fm$^{-3}$. The solid lines indicate our results,
while the dashed lines show the threshold in free space.}
\vspace{3mm}
\end{figure}

It is important, that while the $K^+$-meson energy at zero
momentum increases with the baryon density 
(see FIG.~\ref{laka9}), because of the negative  $\Lambda$ and $\Sigma$ 
potentials the reaction threshold in nuclear matter at  
$\rho_B{<}$1.4$\rho_0$ is shifted below that in free space.

The maximal downward shift of the
reaction threshold in nuclear matter occurs at 
baryon densities around $\rho_B{\simeq}0.6\rho_0$.
This value is the result of  competition between the simple,
linear dependence on density of the vector potentials and the more
complicated, non-linear behaviour of the scalar potentials. (A similar
competition leads to the saturation of the binding energy of normal
nuclear matter in the QMC model.)
Furthermore, the maximum of the downward shift of the
$\pi{+}N{\to}Y{+}K^+$ reaction threshold amounts to roughly 
30~MeV. We also found that at  baryon densities
$\rho_B{>}$0.2~fm$^{-3}$ the strangeness production threshold
in ${\pi}N$ collisions is higher than the free space case.

\section{The $\pi{+}N{\to}\Lambda{+}K$ reaction in nuclear matter}

Now we apply the resonance model to calculate the in-medium
amplitudes. We keep the coupling constant as well as the form
factors at the values found in free 
space. While this assumption certainly cannot be completely correct
in nuclear matter, there are no presently  established 
ways to improve  this
part of our calculation. In principle, since we started from the
reaction amplitude itself, it is possible to include  
in-medium modifications of the coupling constants as well as the 
form factors when reliable calculations of the changes of these
quantities in nuclear matter  become available.
In the following calculations we include the vector and scalar 
potentials for the interacting (initial) nucleons and final kaons and hyperons,
as well as for the intermediate baryonic resonances and $K^*$-meson.

FIG.~\ref{laka1} shows our results for the differential 
cross section for the reaction
$\pi^-{+}p{\to}\Lambda{+}K^0$ 
at the invariant collision energy $\sqrt{s}$=1683 MeV. It is  calculated 
both in  free space (the solid line) and in  nuclear 
matter, at baryon density $\rho_B$=$\rho_0$ (the dashed line) and
$\rho_B$=2$\rho_0$ (the dotted line). For  comparison, we also
show in FIG.~\ref{laka1} the experimental data collected 
in free space~\cite{Knasel,Baker}.  
The important finding is that not only the absolute magnitude, but
also the shape (the dependence on the $\cos \theta$) 
of the $\pi^-{+}p{\to}\Lambda{+}K^0$ differential cross section, 
depends on the baryon density.

\begin{figure}[htb]
\vspace{-3mm}
\psfig{file=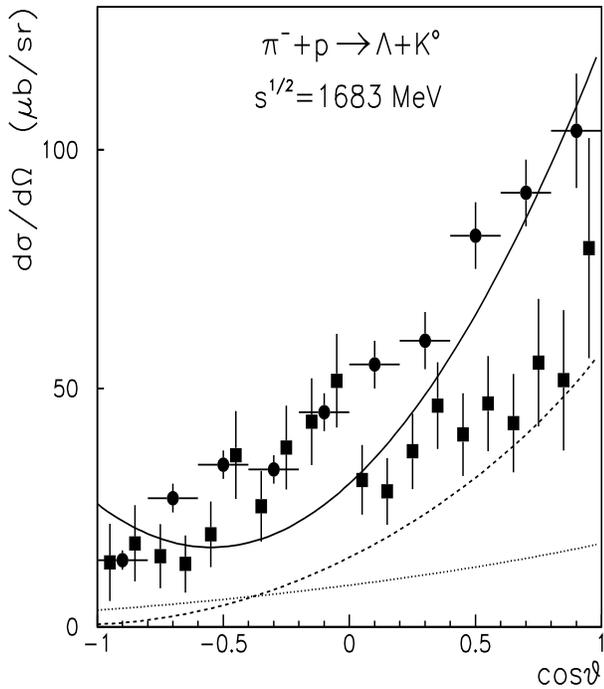,width=9cm,height=11.cm}
\vspace{-3mm}
\caption[]{\label{laka1}The $\pi^-{+}p{\to}\Lambda{+}K^0$ differential
cross section at invariant collision energy $\sqrt{s}$=1683 MeV as
a function of the cosine of the  kaon emission angle in
the center of mass system.
The experimental data are from Ref.~\protect\cite{Knasel} (the squares) and
from Ref.~\protect\cite{Baker} (the circles). The lines show our calculations
in free space (solid) and in nuclear  matter at baryon density
$\rho_B$=$\rho_0$ (dashed) and $\rho_B$=2$\rho_0$ (dotted),
with $\rho_0 = 0.15$ fm$^{-3}$.}
\vspace{5mm}
\end{figure}

One of the simplest ways to construct the in-medium reaction 
cross section is to  take into account  only the in-medium
modification of the flux and phase space factors while leaving amplitude
in matter the same as that in free space, 
without including medium effect~\cite{Pandharipande}.
To shed more light on the problem of how  the reaction amplitude itself
is modified in nuclear matter, we show in  
FIG.~\ref{laka3} the reaction amplitudes squared 
in arbitrary units for
the $\pi^-{+}p{\to}\Lambda{+}K^0$ reaction,  
calculated at $\sqrt{s}$=1.7 GeV and 1.9 GeV, in  
free space (the solid lines), $\rho_B = \rho_0$ (the dashed lines) and 
$\rho_B = 2 \rho_0$ (the dotted lines). 
Our calculation clearly indicates that the $\pi^-{+}p{\to}\Lambda{+}K^0$
reaction amplitude in nuclear matter differs substantially 
{}from that in free space at these energies, 
and that the amplitudes depend on the baryon density.

\begin{figure}[htb]
\vspace{-6mm}
\psfig{file=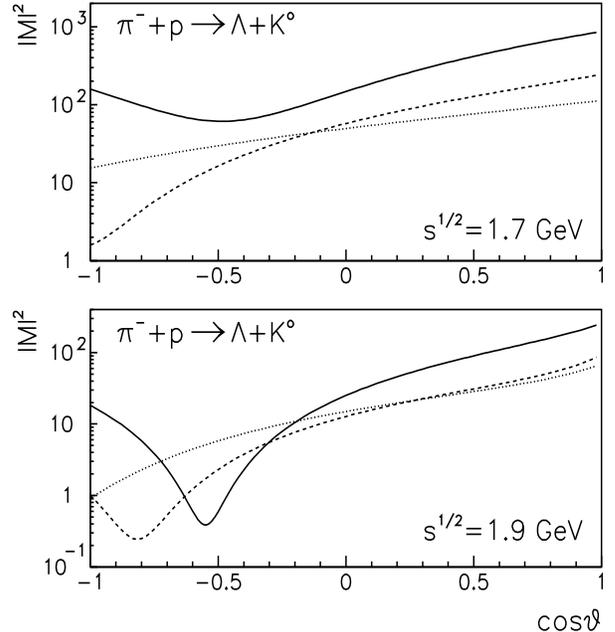,width=9cm,height=10cm}
\vspace{-3mm}
\caption[]{\label{laka3}The (dimensionless) invariant amplitude squared for 
the $\pi^-{+}p{\to}\Lambda{+}K^0$ reaction, 
as a function of $\cos \theta$ (the  $K^+$-meson
emission angle in the center of mass system), calculated for the
invariant collision energies $\sqrt{s}$=1.7 GeV (upper) and
1.9 GeV (lower). The lines show the result for free space (solid)
and for nuclear matter at baryon density $\rho_B$=$\rho_0$ (dashed)
and $\rho_B$=2$\rho_0$ (dotted).}
\vspace{5mm}
\end{figure}

Finally, the energy dependence 
of the total $\pi^-{+}p{\to}\Lambda{+}K^0$
cross section is shown in FIG.~\ref{laka4}, as a function of the
invariant collision energy, $\sqrt{s}$. The experimental data 
in free space are taken from Ref.~\cite{LB}. The calculations for free space 
are in reasonable agreement with the data, as shown by the solid line.
The dashed line in FIG.~\ref{laka4} shows the results obtained for  
nuclear matter at $\rho_B$=$\rho_0$, while the
dotted line is the calculation at $\rho_B$=2$\rho_0$.

Clearly the total $\pi^-{+}p{\to}\Lambda{+}K^0$ reaction cross section 
depends substantially  on the baryon density.
Furthermore, as already discussed in Sec.III,  
the reaction threshold at  
baryon density $\rho_B$=$\rho_0$ is shifted downward as compared to
that in free space, while at $\rho_B$=2$\rho_0$ 
it is shifted upwards. 

\begin{figure}[htb]
\vspace{-2mm}
\psfig{file=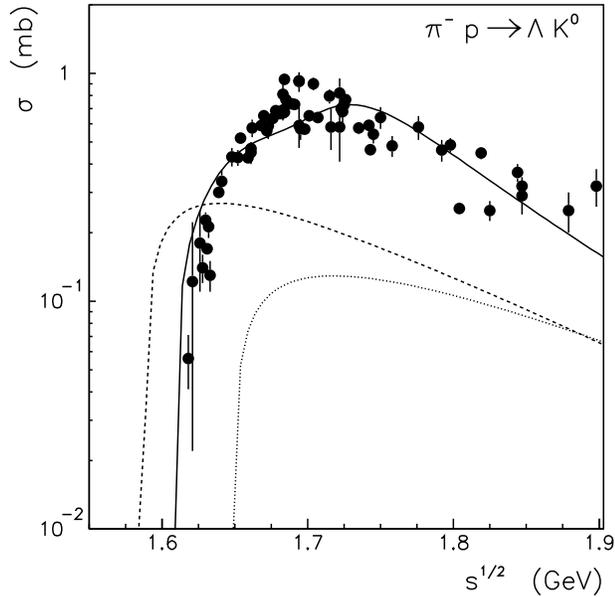,width=9cm,height=9.cm}
\caption[]{\label{laka4}Energy dependence of the 
total cross section, $\pi^-{+}p{\to}\Lambda{+}K^0$, 
as a function of the invariant collision energy, $\sqrt{s}$, 
calculated for different baryon densities.
The data in free space are taken from Ref.~\protect\cite{LB}. 
The lines indicate our results for
free space (solid) and for nuclear matter at baryon density 
$\rho_B$=$\rho_0$ (dashed) and $\rho_B$=2$\rho_0$ (dotted). (Only the
solid curve should  be compared directly with the data.)}
\vspace{5mm}
\end{figure}

Obviously,  heavy ion collisions probe a range of baryon densities from  
$\rho_B$=0 up to several times 
normal nuclear matter density, $\rho_0$. The calculation of the time
and spatial dependence of the baryon density distribution is 
a vital aspect of  dynamical heavy
ion simulations. However, a first estimate 
of the density averaged total $\pi^-{+}p{\to}\Lambda{+}K^0$ 
cross section can be gained from FIG.~\ref{laka4}. 
Of course, the data is only available in free space and
should only be directly compared with the solid curve. Nevertheless, it
is suggestive for the problem of in-medium production to note that a
crude average of the in-medium cross sections over 
the range 0${<}\rho_B{<}2\rho_0$ would be quite close to the free space
cross section at energies around the free space 
threshold. This seems to provide a reasonable
explanation of why the heavy 
ion calculations including~\cite{Cassing3,Senger3,Li3}  the 
$\pi{+}p{\to}\Lambda{+}K$  cross section in 
free space, that is without a repulsive kaon potential, can  
reproduce the data~\cite{Schroter,Senger1,Barth,Laue,Senger2}. 
More quantitative calculation and discussion of this effect will be given 
in Sec.~VI.

\section{The $\pi{+}N{\to}\Sigma{+}K$ reaction in nuclear matter}
The $\pi{+}N{\to}\Sigma{+}K$ reaction involves different dynamics
in comparison with the $\pi{+}p{\to}\Lambda{+}K$ reaction, because
the reaction involves the different intermediate baryonic
resonances. For instance, although the $N(1650)$ resonance couples
to ${\Lambda}N$ channel, it does not couple to the ${\Sigma}N$ state. 
Moreover, the $\Delta(1920)$ resonance couples to  
${\Sigma}N$ channel, but does not couple to the ${\Lambda}N$ channel 
in our model. 
{}For this reason, the dependence on the baryon density of the  
reaction in nuclear matter, $\pi{+}N{\to}\Sigma{+}K$, 
is quite different from that of  
the $\pi{+}N{\to}\Lambda{+}K$.

\begin{figure}[htb]
\vspace{-2mm}
\psfig{file=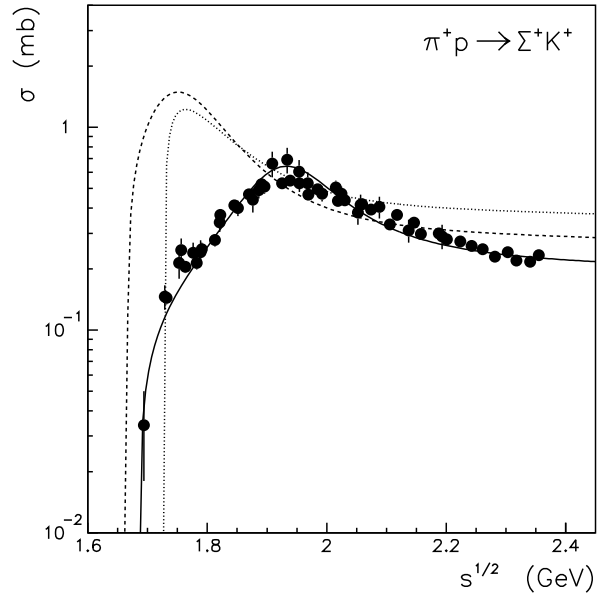,width=9cm,height=9cm}
\caption[]{\label{laka5}Energy dependence of the
total cross section, $\pi^+{+}p{\to}\Sigma^+{+}K^+$, 
as a function of the invariant collision energy, $\sqrt{s}$, 
calculated for different baryon densities.
The data in free space are taken from Ref.~\protect\cite{LB}. The 
lines indicate our calculations for
free space (solid) and for nuclear matter at baryon density 
$\rho_B$=$\rho_0$ (dashed) and $\rho_B$=2$\rho_0$ (dotted).}
\vspace{5mm}
\end{figure}

In FIG.~\ref{laka5} we show the energy dependence of the 
total cross section, $\pi^+{+}p{\to}\Sigma^+{+}K^+$,  
as a function of the invariant collision energy, 
$\sqrt{s}$. The experimental data in free space are taken from 
Ref.~\cite{LB}. The free space data are well 
reproduced by the calculations in free space, as shown in 
FIG.~\ref{laka5} by the solid line. The dashed line indicates the
results obtained for nuclear matter at baryon density
$\rho_B$=$\rho_0$, while the dotted line shows the result at 
$\rho_B$=2$\rho_0$. 

Again, as already discussed in Sec.~III, the 
density dependence of the hadron masses and the vector 
potentials leads to a
shift of the reaction thresholds
in nuclear matter. Because of the density dependence of the
$\Sigma$-hyperon potential, the threshold at normal nuclear matter 
density ($\rho_B$=$\rho_0$) is shifted downwards 
compared with that in free space.
At $\rho_B$=2$\rho_0$ the ${\Sigma}K$ reaction threshold
is shifted upwards relative to the threshold in free space.
Moreover, the magnitude of the $\pi^+{+}p{\to}\Sigma^+{+}K^+$ cross section
depends much more strongly on the density than the
$\pi^-{+}p{\to}\Lambda{+}K^0$ reaction. 

FIGs.~\ref{laka6} and \ref{laka7} show the energy dependence of the 
total cross sections
for the $\pi^-{+}p{\to}\Sigma^0{+}K^0$ and $\pi^-{+}p{\to}\Sigma^-{+}K^+$
reactions, respectively. The data in free space~\cite{LB} 
are well reproduced 
with the calculations in free space, which are indicated by the solid lines.
The cross sections calculated for the nuclear matter, 
except for $\pi^-{+}p{\to}\Sigma^0{+}K^0$ at $\rho_B = 2 \rho_0$, are 
substantially enhanced in comparison with those in free space, 
at energies above near the in-medium reaction 
thresholds.

\begin{figure}[htb]
\vspace{-2mm}
\psfig{file=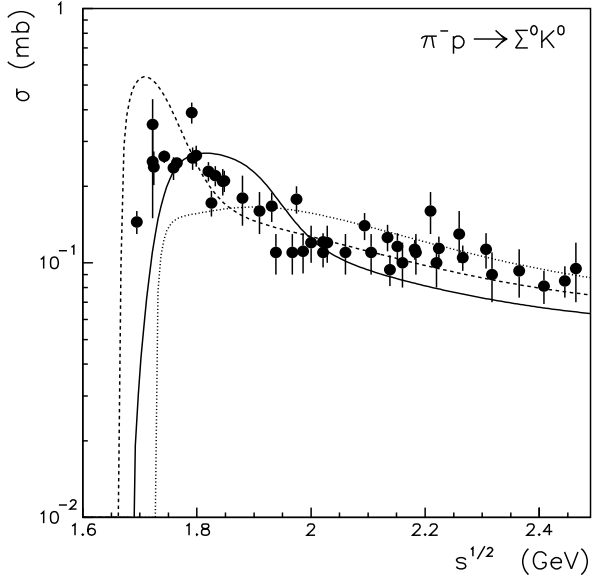,width=9cm,height=8.68cm}
\caption[]{\label{laka6}Energy dependence of the
total cross section, $\pi^-{+}p{\to}\Sigma^0{+}K^0$,
as a function of the invariant collision energy, $\sqrt{s}$, 
calculated for different baryon densities.
The data in free space are taken from Ref.~\protect\cite{LB}. The 
lines indicate our calculations for
free space (solid) and for nuclear matter at baryon density 
$\rho_B$=$\rho_0$ (dashed) and $\rho_B$=2$\rho_0$ (dotted).}
\vspace{5mm}
\end{figure}

\begin{figure}[htb]
\vspace{-12mm}
\psfig{file=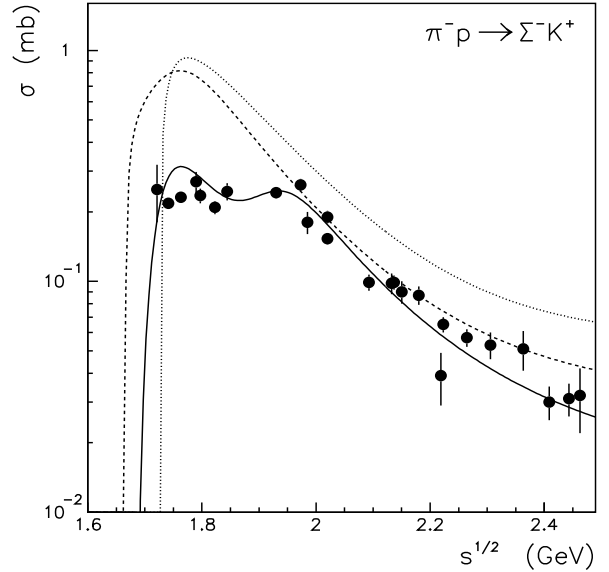,width=9cm,height=8.68cm}
\caption[]{\label{laka7}Energy dependence of the
total cross section, $\pi^-{+}p{\to}\Sigma^-{+}K^+$,
as a function of the invariant collision energy, $\sqrt{s}$,
calculated for different baryon densities.
The data in free space are taken from Ref.~\protect\cite{LB}. The 
lines indicate our calculations for
free space (solid) and for nuclear matter at baryon density 
$\rho_B$=$\rho_0$ (dashed) and $\rho_B$=2$\rho_0$ (dotted).}
\vspace{4mm}
\end{figure}

\section{Impact on heavy ion studies}
It is expected that in relativistic heavy ion 
collisions at SIS energies  nuclear matter can be compressed up to 
baryonic densities of order $\rho_B{\simeq}{3}\rho_0$~\cite{Senger3}. 
The baryon density
$\rho_B$ available in heavy ion collisions evolves with the interaction 
time, $t$, and is given by the dynamics of the heavy ion collision.
Moreover, the density is large in the very center of the
collision. In the following estimates we investigate
the density dependence of the production cross section for central 
central heavy ion collisions. However, it should be remembered 
that at the edges, 
where most particles are expected to be located,   
the density dependence of the strangeness production
mechanism is not strong compared to that of the central 
zone of the collision.

To calculate the $K^+$-meson production cross section averaged
over the available density distribution we adopt the 
density profile function obtained by dynamical 
simulations~\cite{Hombach} for $Au{+}Au$ collisions at 
2 AGeV and at impact 
parameter $b{=}0$. This can be parametrized as 
\begin{equation}
\label{time1}
\rho_B (t) = \rho_{max} \, \exp 
\left( \frac{[\, t\,- \,{\bar t} \, \, ]^2}{{\Delta t}^2}
\right),
\label{dprofile}
\end{equation} 
where the parameters, $\rho_{max}$=3$\rho_0$, 
${\bar t}$=13~fm and ${\Delta t}$=6.7~fm, were fitted to the heavy
ion calculations~\cite{Hombach}.

\begin{figure}[htb]
\vspace{-2mm}
\psfig{file=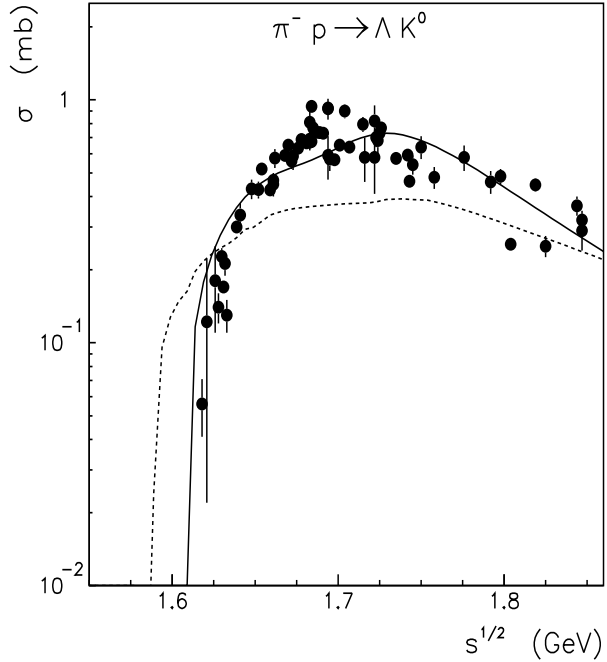,width=9cm,height=10.cm}
\caption[]{\label{laka4a}Energy dependence of the
total cross section, $\pi^-{+}p{\to}\Lambda{+}K^0$, 
as a function of the invariant collision energy, $\sqrt{s}$.
The data in free space are taken from Ref.~\cite{LB}. The 
solid line indicates our calculation for
free space. The dashed line shows the cross section calculated 
by averaging over the density function profile~\protect\cite{Hombach}
given by the time evolution obtained for 
$Au{+}Au$ collisions at 2 AGeV (see Eq.~(\protect\ref{dprofile})).}
\vspace{5mm}
\end{figure}

The total cross section for the $\pi^-{+}p{\to}\Lambda{+}K^0$ reaction 
integrated over the time range $5{\le}t{\le}23$~fm and weighted by 
the time dependent density profile given in Eq.~(\ref{dprofile}),  
is shown by 
the dashed line in FIG.~\ref{laka4a}. The 
limits of the $t$ integration were taken from the simulations of the
$Au{+}Au$ collision time evolution in Ref.~\cite{Hombach}. The circles and 
solid line in FIG.~\ref{laka4a} show the experimental data 
in free space~\cite{LB}
and the calculations in free space, respectively. 

One can see that the total cross section averaged over 
the collision time (time dependent density profile) for 
the $\pi^-{+}p{\to}\Lambda{+}K^0$ reaction 
is quite close to the result given in
free space integrated up to at energies above the production threshold, 
up to $\sqrt{s} \simeq$ 1.7 GeV.
That the results shown in FIG.~\ref{laka4a} actually explain why the 
heavy ion calculations with the free space kaon production cross
section might quite reasonably reproduce the experimental data, 
will be discussed more quantitatively below.

As a matter of fact, the total cross section averaged over 
the time dependent density profile,  
shown by the dashed line in FIG.~\ref{laka4a}, 
should additionally be averaged over the
invariant collision energy distribution available in heavy
ion reactions. The number of meson-baryon collisions, $N_{mB}$, 
for the central $Au{+}Au$ collisions at 2 AGeV is given 
in Ref.~\cite{CBJ} as a function of the invariant collision energy, 
$\sqrt{s}$. It can be parametrized for $\sqrt{s}{>}$1~GeV as
\begin{equation}
\frac{dN_{mB}}{d\sqrt{s}} = N_0 \, \exp{\left( \frac{[\, 
\sqrt{s} \,- \,{\sqrt{s_0}} \, \, ]^2}{[{\Delta \sqrt{s}}]^2}
\right)},
\label{donsi}
\end{equation}
where the normalization factor $N_0$=6$\times$10$^4$ GeV$^{-1}$, 
while $\sqrt{s_0}$=1 GeV
and $\Delta \sqrt{s}$=0.63~GeV. Note that, at SIS
energies $N_{mB}$ is almost entirely given by the pion-nucleon
interactions, and heavy meson and baryon collisions contribute 
only to the high energy tail of the 
distribution in Eq.~(\ref{donsi}) -- with quite small densities~\cite{CBJ}.  
{}Finally, if we also average the 
calculated, in-medium, total cross section for 
$\pi^-{+}p{\to}\Lambda{+}K^0$, shown by the dashed line
in FIG.~\ref{laka4a}, over the available energy distribution given
in Eq.~(\ref{donsi}), we obtain an average total 
kaon production cross section of 
${<}K{>}$=65~$\mu$b for 
central $Au{+}Au$ collisions at 2 AGeV. 
This result is  indeed 
compatible with the calculations using the free space total cross section 
of the $\pi^-{+}p{\to}\Lambda{+}K^0$ reaction, 
which provide an average total kaon production 
cross section of ${<}K{>}$=71~$\mu$b
for central $Au{+}Au$ collisions at 2 AGeV. 
Note that the inclusion of even a slight modification of the 
$K^+$ mass because of the nuclear medium 
(without the corresponding changes introduced here) 
leads to a substantial reduction of the inclusive $K^+$ spectra
(by as much as a factor of 2 or 3), 
compared to that calculated using the free space properties 
for the relevant hadrons~\cite{Bratkovskaya}.


We stress that at SIS energies  reaction channels with 
a $\Sigma$-hyperon in the final state play a minor role, 
because of the upper limit of the energy available in the 
collisions.  As was illustrated by FIG.~\ref{laka10},
the downwardly shifted $\pi{+}N{\to}\Sigma{+}K$ reaction threshold 
at small densities is still quite high 
compared to that for the reaction with 
the $\Lambda$-hyperon in the
final state.

\section{Summary}
We have calculated the in-medium modification of  kaon production
in pion-nucleon collisions in nuclear matter using the 
resonance model developed in 
Refs.~\cite{Tsushima2,Tsushima3,Tsushima3a,Tsushima4}. To evaluate
the in-medium $K$-meson production for the  reaction channels
with $\Lambda$ and $\Sigma$ hyperons in the final states, the density
dependent potentials for the initial, final and intermediate
mesons and baryons (resonances)
were introduced to the resonance model amplitudes.
The vector and scalar potentials were calculated within the
quark-meson coupling
model~\cite{Tsushima1,Guichon,Tsushimak1,Tsushimak2,Tsushimak3,Tsushimak4,Tsushimak5,Saitok1,Saitok2,Tony,Hyperon1,Hyperon2}.
The $\pi{+}N{\to}\Lambda{+}K$ and  $\pi{+}N{\to}\Sigma{+}K$ 
cross sections were calculated for different
baryon densities of the nuclear matter. We found, that not only
are the initial flux and the final phase space of the reactions  
modified in baryonic matter, but the reaction amplitudes 
themselves are also modified.

It was shown, that the total $\pi{+}N{\to}\Lambda{+}K$ and  
$\pi{+}N{\to}\Sigma{+}K$ cross sections depend substantially on the 
baryonic density. Furthermore, the reaction thresholds and the 
absolute magnitudes
as well as the dependence of the production cross section
on the invariant collision energy, $\sqrt{s}$, all vary strongly with the
density of the nuclear matter. 

To evaluate the impact of our microscopic calculations on the
heavy ion results, we averaged the kaon production cross section
over the baryon density profile, which depends on the evolution
time of the heavy ion collision. Furthermore, in order to compare 
with the experimental data more quantitatively, we calculated 
the effective total kaon production cross section 
by averaging over the invariant collision energy 
distributions available in heavy ion reactions. We found that
at low collision energies, 
the density or time averaged $K^+$-meson production total 
cross section, calculated using 
the in-medium properties for the $K^+$ meson, hyperons and relevant 
hadrons, is very close to that calculated using the total cross section 
given in free space.

Thus, our results provide an explanation of why the 
analyses~\cite{Li2,Cassing3,Senger3,Li3} of  
available data on $K^+$ production from heavy ion collisions at SIS 
energies~\cite{Barth,Laue,Senger2,Senger3} for the 
$K^+$ spectra, can be reasonably described when one neglects
any in-medium modification of the kaon and hadronic properties   
-- i.e., adopting  
the $K^+$-meson production cross section given in free space.

In conclusion, our present study shows that if one accounts for 
the in-medium modification of the production amplitude   
(i.e., the in-medium properties of the $K^+$-meson and hadrons) correctly,  
it is possible to understand $K^+$ production data in heavy ion collisions 
at SIS energies, even if the $K^+$-meson feels the theoretically expected,  
repulsive mean field potential. The apparent failure to explain  
the $K^+$ production data if one includes the purely kinematic 
effects of the in-medium modification 
of the $K^+$-meson and hadrons, appears to be a consequence of the
omission of these effects on the reaction amplitudes.

\acknowledgements
A.S. would like to acknowledge the warm hospitality at the CSSM during
his visit. This work was supported by the
Australian Research Council and the Forschungszentrum J\"ulich.

\end{document}